\documentstyle[psfig,epsf,epsfig,12pt]{article}
\newcommand{\CPC}[3]{Comput. Phys.\ Commun.\ {\bf#1} (#3) #2}

\newcommand{\NP}[3]{Nucl.\ Phys.\ {\bf#1} (#3) #2}
\newcommand{\PL}[3]{Phys.\ Lett.\ {\bf#1} (#3) #2}
\newcommand{\PR}[3]{Phys.\ Rev.\ {\bf#1} (#3) #2}
\newcommand{\PRL}[3]{Phys.\ Rev.\ Lett.\ {\bf#1} (#3) #2}

\newcommand{\PRP}[3]{Phys.\ Rep.\ {\bf#1} (#3) #2}
\newcommand{\ZP}[3]{Z.\ Phys.\ {\bf#1} (#3) #2}

\newcommand{\NIM}[3]{Nucl.\ Instr.\ and Meth.\ {\bf#1} (#3) #2}
\newcommand{\xbj}{{x_{{\rm BJ}}}}
\newcommand{\yjb}{{y_{{\rm JB}}}}
\newcommand{\etamax}{{\eta_{{\rm max}}}}
\newcommand{\mx}{{M_X}}
\newcommand{\xl}{{x_{{\rm L}}}}
\newcommand{\ptr}{{p_{{\rm T}}}}
\newcommand{\runc}{{r_{\rm unc}}}
\newcommand{\rbar}{{\overline{r}_{\rm unc}}}
\newcommand{\pomeron}{{{\rm I\!P}}}
\newcommand{\lap}{{\stackrel{_{\scriptstyle <}}{_{\scriptstyle\sim}}}}
\newcommand{\gap}{{\stackrel{_{\scriptstyle >}}{_{\scriptstyle\sim}}}}

\def\mc#1 {\multicolumn{1}{|c|}{#1}}

\begin{document}           

\title{Observation of Events with an Energetic Forward
Neutron in Deep Inelastic Scattering at HERA}
\author{ZEUS Collaboration}
\date{  }
\maketitle


\vspace{5 cm}
\begin{abstract}
In deep inelastic neutral current scattering of positrons
and protons at the center of mass energy of 300 GeV, we observe, 
with the ZEUS detector,
events with a high energy neutron produced at very small scattering
angles with respect to the proton direction.
The events 
constitute a fixed fraction of the deep inelastic, neutral
current event sample independent of Bjorken $x$ and $Q^2$
in the range $ 3 \cdot 10^{-4} < \xbj < 6\cdot 10^{-3}$
and $10 < Q^2 < 100$~GeV$^2$.
\end{abstract}
\setcounter{page}{0}
\thispagestyle{empty}   
\newpage
\vspace*{0.5 cm}
\section{Introduction}
 
The general features of the hadronic final state in deep
inelastic lepton nucleon scattering (DIS) are well
described by models inspired by Quantum Chromodynamics (QCD).
In these models the struck quark and the colored proton remnant
evolve into a system of partons which fragments into hadrons.
Many of these models neglect peripheral processes, which are 
characterized by leading baryons.



A recent example of peripheral processes is the observation
by ZEUS~\cite{z_rap_gap} and H1~\cite{h_rap_gap} of DIS
events with large rapidity gaps.
These events are distinguished by
the absence of color flow between the final state baryonic system and the
fragments of the virtual photon, and they have been interpreted
as arising from diffraction.
In the language of Regge trajectories, a pomeron $\pomeron$,
with the quantum numbers of the vacuum, is exchanged between the proton and
the virtual photon.

Another example is 
provided by 
meson exchange~\cite{sullivan,zoller,levman,koepf,kopeliovich,ingelman},
which plays
a major role in peripheral hadronic scattering.
In this process, the incoming proton fluctuates into a baryon and a meson.
At HERA energies, the lifetime of this state can be sufficiently long  
that the lepton may interact with the meson.
In $p\rightarrow p$ 
transitions the exchange of neutral mesons occurs together with diffractive
scattering. These contributions may be separable by measuring the 
proton momentum distribution.
On the other hand,   
$p\rightarrow n$ transitions  signal events where
charged 
meson exchange could dominate~\cite{bishari,holtmann}, regardless of the
neutron momentum.
The pion, being the lightest meson, may provide the largest contribution
to the cross section.
Isolation of the one pion exchange contribution would 
provide the opportunity to study virtual gamma pion interactions
and thereby determine the structure function of the
pion.

In order to study these issues we have installed a hadronic calorimeter to
detect high energy forward going neutrons produced in DIS 
($ep\rightarrow en+$anything) at HERA.
This paper reports the first observation of such events,
showing clear evidence of
sizeable leading neutron production.

\section{Experimental setup}

The data were collected with the ZEUS detector during 
1994 while HERA collided 153 $ep$ bunches of
27.5 GeV positrons and 820 GeV protons.
In addition, 15 unpaired bunches of
positrons and 17 unpaired bunches of protons
circulated, 
permitting a measurement of beam associated
backgrounds.
The data sample used in this analysis corresponds to an integrated luminosity
of 2.1 pb$^{-1}$. 
 
\begin{figure}
\centerline{\psfig{file=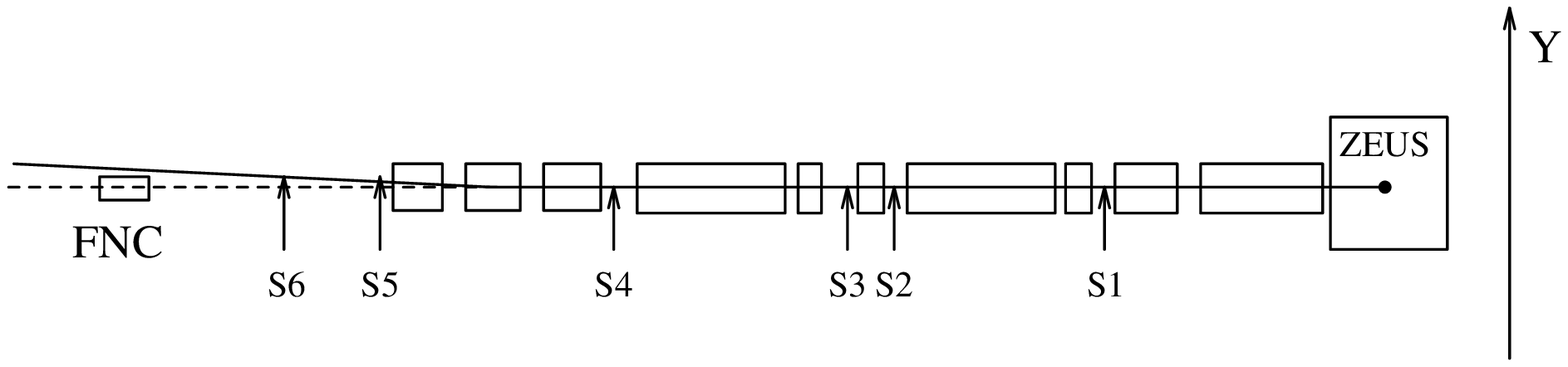,width=12cm}}
\centerline{(a)}
\vspace{1cm}
\centerline{\psfig{file=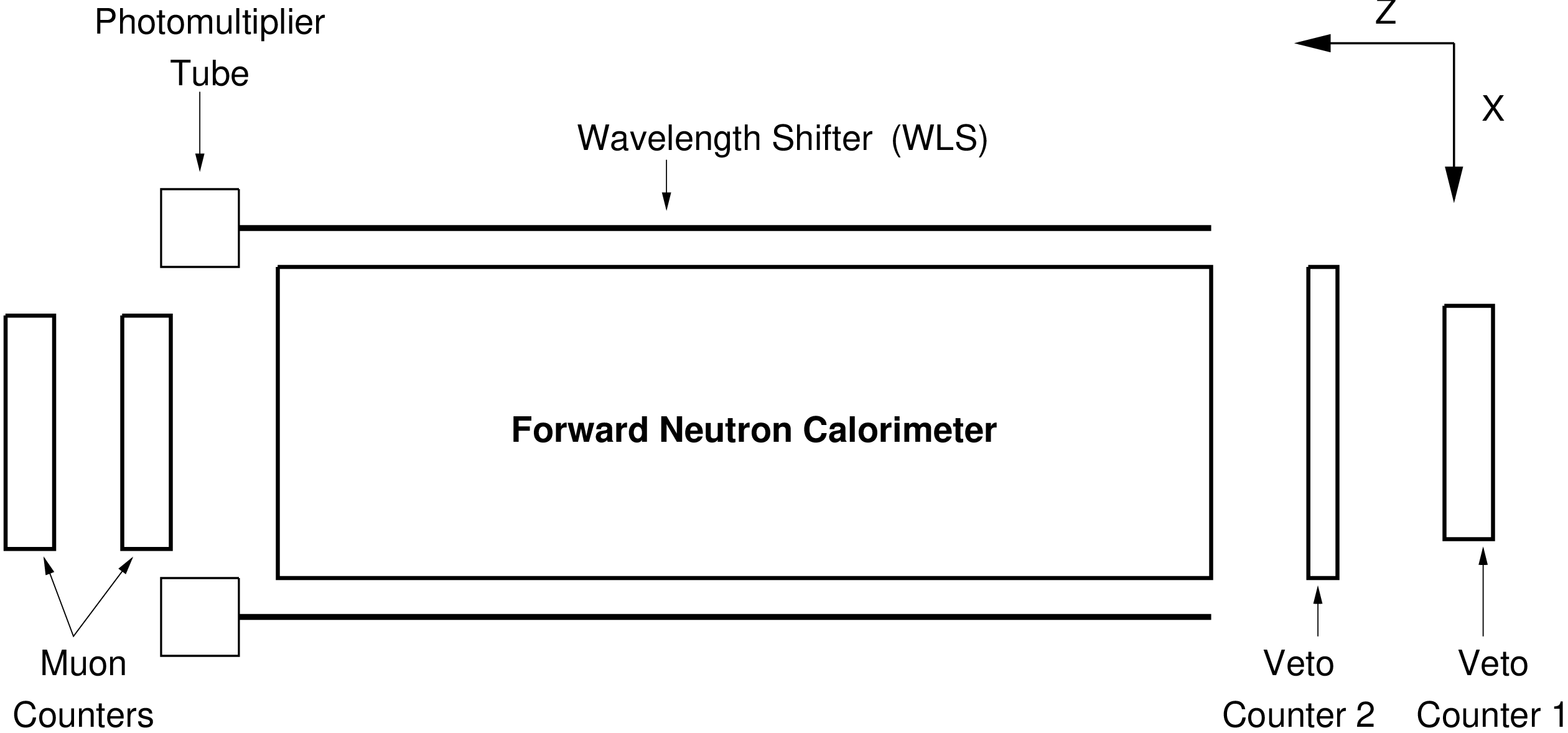,width=5cm,angle=-90}}
\centerline{(b)}
\vspace{1cm}
\begin{minipage}[t]{5cm}
\centerline{\hspace{2cm}\psfig{file=fig1c.eps,width=3cm,angle=90}}
\centerline{\hspace{2cm}(c)}
\end{minipage}\hspace{3cm} 
\begin{minipage}[t]{5cm}
\centerline{\psfig{file=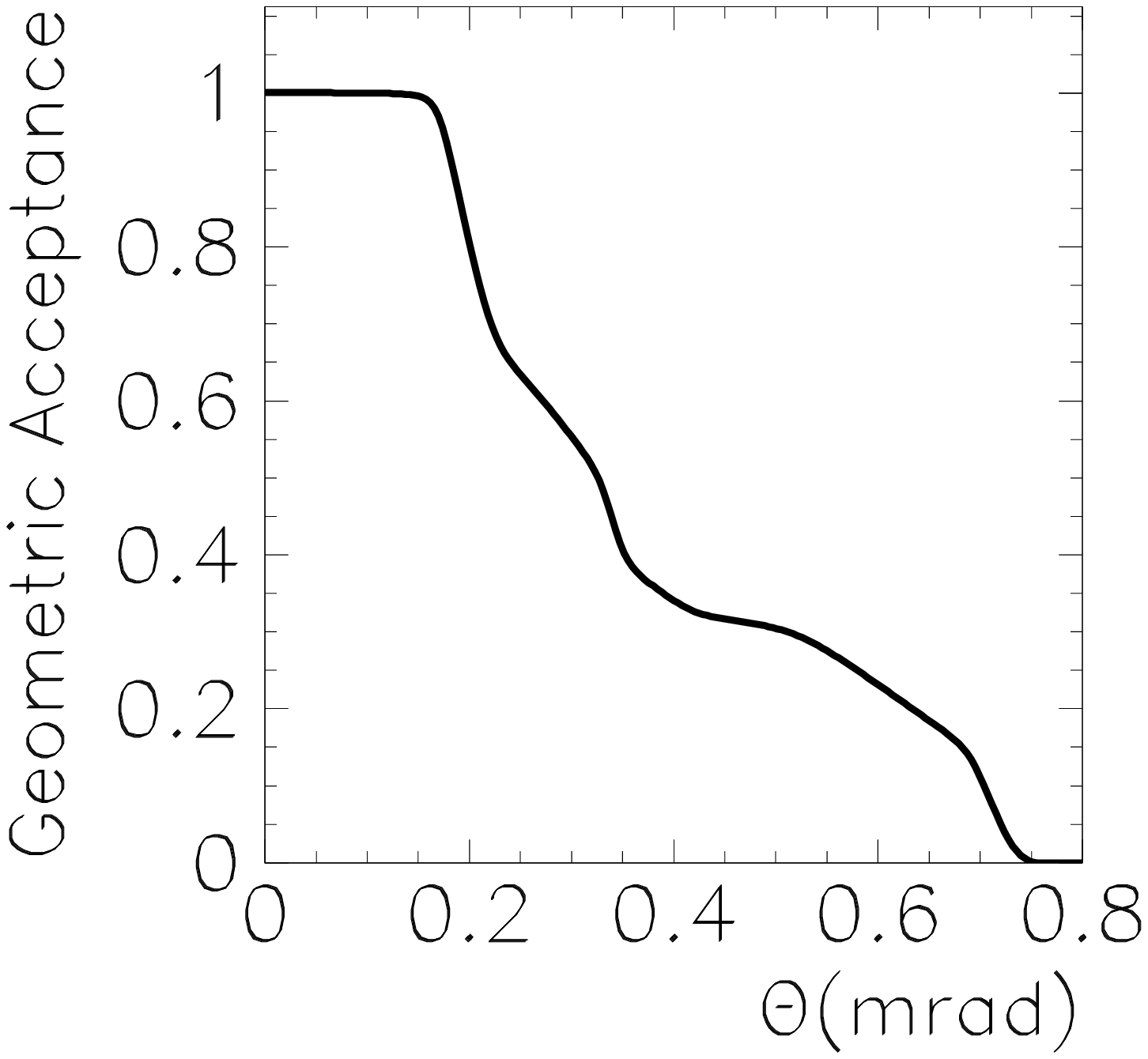,width=5cm}}
\centerline{\hspace{0.5cm}(d)}
\end{minipage}
\vspace{1cm}
\caption{(a) Schematic layout of the proton beam line viewed
from the side with FNC II (at $Z = 101$ m) below the beam pipe and
downstream of LPS stations S1-S6.
(b) Schematic drawing of FNC II viewed from the top.
(c) Front view of FNC II showing the segmentation into three towers,
and the projected region of geometric aperture allowed
by the HERA magnets.
The cross indicates the position of the zero degree line. (d) The
geometric acceptance as a function of polar angle
(scattering angle), integrated over azimuth.}
\label{calfig}
\end{figure}

The present analysis makes use of a test Forward Neutron Calorimeter
(FNC II)~\cite{prc} 
installed at the beginning of 1994 in the HERA tunnel at
$\theta = 0$ degrees, $Z = 101$ m, downstream of the interaction 
point\footnote{The ZEUS coordinate system is defined 
as right handed with
the $Z$ axis pointing in the proton beam direction and the $X$
axis horizontal, pointing towards the center of HERA.}.
The layout of the beam line and calorimeter
is shown schematically in Fig.~\ref{calfig}.
FNC II, located after the final station of the
ZEUS Leading Proton Spectrometer (LPS), 
was an enlarged and improved version of the original
test Forward Neutron Calorimeter (FNC I) which operated in
1993. The design, construction and calibration
of FNC II was similar to FNC I~\cite{bhadra,brkic}.
Both devices were iron-scintillator sandwich calorimeters
read out with wavelength shifter light guides coupled
to photomultiplier tubes (PMT).
The unit cell consisted of 10 cm of iron followed by 0.5 cm of 
SCSN-38 scintillator. FNC II contained 17 unit cells
comprising a total depth of 10 interaction lengths. It was
40 cm wide and 30 cm high, divided vertically
into three 10 cm towers read out on both sides. 
There was no longitudinal subdivision in the readout.

The neutron calorimeter was situated downstream of the HERA dipoles
which bend the 820 GeV proton beam upwards.
Charged particles originating at the interaction point
were swept away from FNC II.
The aperture of the HERA magnets in
front of FNC II
limited the geometric acceptance as shown 
in Figs.~\ref{calfig}(c) and (d).   
Between these magnets and  FNC~II the neutrons 
encountered inactive material, the thickness of 
which varied between one and two interaction
lengths.
Two scintillation veto counters preceded the calorimeter: 
one 30\,x\,25\,x\,5~cm$^3$, and one 40\,x\,30\,x\,1~cm$^3$. 
These counters were used offline to identify
charged particles and thereby reject
neutrons which interacted in the inactive material
in front of FNC II.
The calorimeter was followed by two scintillation
counters,
which were used
in coincidence with the front counters 
to identify beam halo muons. 
The response of the counters to
minimum ionizing particles was determined with these muons.

Energy deposits in  
FNC II were read out using a system identical to that
of the ZEUS uranium
scintillator calorimeter (CAL).
In addition the rate of signals exceeding a threshold 
of 250 GeV  was recorded.
The accumulated counts
provide the average counting 
rate of FNC II for each run.  
 
The other components of ZEUS have been described
elsewhere~\cite{b:Detector}.
The CAL, the central
tracking detectors (CTD,VXD), the small angle rear tracking
detector (SRTD) which is a scintillator hodoscope in front of
the rear calorimeter close to the beam pipe, and
the luminosity monitor (LUMI) are the
main components used for the analysis
of DIS events~\cite{z_shift}.

\section{Kinematics of deep inelastic events }
 

In the present analysis the two particle inclusive reaction
$ep \rightarrow en+$anything is compared with the single
particle inclusive reaction $ep \rightarrow e+$anything.
In both cases the scattered positron and part of the hadronic
system, denoted by X, were detected in CAL. Energetic forward
neutrons were detected in FNC~II. The two particle inclusive events
are specified by
four independent kinematical variables: any two of
$\xbj$, $Q^2$, $y$, and $W$ for the scattered lepton; 
and any two of $\xl$, $\ptr$, and $t$ 
for the leading baryon (see below). 

Diagrams
for one and two particle inclusive $ep$ scattering are shown
in Fig.~\ref{plot1}(a) and (b).   
The conventional DIS kinematical variables describe the
scattered positron: $Q^2$, the negative of the squared four--momentum
transfer carried by the virtual photon $\gamma^*$,
\[ Q^2\equiv -q^2 = -(k-k')^2,\]
where $k$ and $k'$ are the four--momentum vectors of the initial and
final state positron respectively; $y$,
the energy transfer to the hadronic final state
\[ y \equiv\frac {q \cdot P}{k \cdot P} \ , \] 
where $P$ is the four--momentum
vector of the incoming proton; $\xbj$, the Bjorken variable 
\[ \xbj\equiv\frac{Q^2}{2q \cdot P} = \frac{Q^2}{y s}\ , \] 
where $s$ is the
center-of-mass 
(c.m.) energy squared of the $ep$ system; and $W$, the 
c.m. energy of the $\gamma^*p$ system,
\[W^2\equiv(q+P)^2=\frac{Q^2(1-\xbj)}{\xbj}+{M_p}^2, \] 
where $M_p$ is the mass of the proton. 
 
The ``double angle method'' was used to determine
$\xbj$ and $Q^2$~\cite{bentvelsen}. In this method,
event variables are derived 
from the scattering angle
of the positron and the scattering angle $\gamma_H$ of 
the struck (massless) quark.
The latter angle is determined
from the hadronic energy flow measured in the main ZEUS detector,
\[\cos\gamma_H=\frac{(\sum_i p_X)^2 +(\sum_i
p_Y)^2-(\sum_i(E-p_Z))^2}{(\sum_i p_X)^2 +(\sum_i
p_Y)^2+(\sum_i(E-p_Z))^2}\ , \] 
where the sums run over all CAL
cells $i$, excluding those assigned to the scattered
positron, and ${\rm\bf p}=(p_X, p_Y, p_Z)$ is the momentum vector assigned to
each cell of energy $E$.  The cell angles are calculated from the
geometric center of the cell and the vertex position of the event.
Final state particles produced close to the direction of the proton
beam give a negligible contribution to $\cos\gamma_H$, since these particles have
$(E-p_Z) \simeq 0 $.
 
In the double angle method, in order that the hadronic system
be well measured, it is necessary to require a minimum 
hadronic energy in the CAL away from the beam pipe.
A suitable quantity for this purpose is the hadronic
estimator of the variable $y$, defined by
\[
\yjb \equiv \frac{\sum_i(E-p_Z)}{2E_e},
\]
where $E_e$ is the electron beam energy.

The two independent kinematical variables describing the neutron
tagged by FNC~II are 
taken to be its energy $E_n$ and transverse momentum $\ptr$.
These quantities are related to the four-momentum
transfer squared between the proton and the neutron, $t$, by
\[
t \simeq -\frac{\ptr^2}{\xl}-\frac{(1-\xl)}{\xl}\left(M^2_n-\xl M^2_p\right),
\]
where $M_n$ is the mass of the neutron and 
$\xl \equiv E_n/E_p$, where 
$E_p$ is the proton beam energy.
The geometry of FNC II and the HERA beam line
limited
the angular acceptance of the scattered neutron
to $\theta \lap 0.75$ mrad, and the threshold on
energy deposits in FNC II  restricted $\xl$ to
$\xl > 0.3$.

The invariant mass of the hadronic system detected in the calorimeter,
$M_X$, can be determined from the cell information in CAL;
an approach similar to the double angle method is applied 
to calculate $M_X$.
Given the energy, $E_H$, the momentum, $p_H$, 
and the polar angle, $\theta_H$, of the hadronic system
observed in the detector, the following formulae
determine $M_X$: $\cos \theta_H =\sum_i p_Z /|\sum_i {\rm\bf p}|$,
where the sum runs over all calorimeter cells $i$, excluding those
assigned to the positron,
$p_H^2 = Q^2(1-y)/\sin^2\theta_H$,
$E_H= 2E_e y+p_H \cos\theta_H$,
$M_X=\sqrt{E_H^2-p_H^2}$.
 
The identification of neutral current deep
inelastic events uses the quantity $\delta$ defined by 
\[ \delta \equiv \sum_i (E-p_Z)\ , \] 
where the sum runs over all CAL cells
$i$.  For fully contained neutral current DIS events, and neglecting
CAL resolution effects and initial state radiation, $\delta = 2E_e$.

We also use the variable $\etamax$ which is defined as the
pseudorapidity,
\[
\eta \equiv -\ln \tan \left( \theta /2 \right),
\]
of the calorimeter cluster
with energy greater than 400 MeV
closest to the proton beam direction.

\section{Monte Carlo simulation and studies}
\label{mc}

The response of FNC II was modeled by a Monte Carlo (MC) 
simulation using
the GEANT program~\cite{geant}. The model was inserted into
the full simulation of the ZEUS detector and beam line.
For neutrons incident on the face of the
calorimeter the predicted energy
resolution is approximately $\sigma (E_n) = 2.0\sqrt E_n$,
with $E_n$ in GeV.
The predicted energy response of the calorimeter 
is linear to better than 5\%. 

To aid the study of energetic neutron production both
in beam gas collisions and in DIS,
a Monte Carlo generator was written
for one pion exchange,
which gives a cross
section proportional to
$|t|\cdot (1-\xl)^{1-2\alpha_{\pi}(t)}/\alpha^2_{\pi}(t)$
 (see, for example,~\cite{bishari}),
where $\alpha_{\pi}(t)=\alpha'_{\pi}(t-m^2_{\pi})$
is the pion trajectory and $\alpha'_{\pi}=1~$GeV$^{-2}$. 
The code uses, as a framework, the HERWIG program~\cite{herwig}. 
Absorptive corrections to one pion exchange have been
widely discussed (see, for example,
\cite{williams,gotsman,schrempp}). To estimate such effects 
a simple prescription which replaces $|t|$ by $|t| + m^{2}_{\pi}$ in
the numerator of the above expression was used. 
In addition to the one pion exchange model, 
the standard QCD inspired DIS models 
ARIADNE \cite{ariadne}, HERWIG, and MEPS \cite{meps} were used to predict
the forward neutron production.

To compare data with the expectations of 
all these models,
the MC events produced by the generators were fed through
the simulation of the ZEUS detector.

\section{Calibration and acceptance of FNC II}

The relative gains of the PMTs were determined by scanning
each tower with a $^{60}$Co gamma source using the
procedure developed for the ZEUS CAL~\cite{co60}. This was done
at the end of the data taking period.
Beam gas data taken in HERA were used for calibration.
These data were obtained
after the proton beam was accelerated to 820 GeV, but before positrons
were injected. To reject events where the neutrons had showered
in material upstream of FNC II, events were considered only when
the energy deposited in the veto counters was below that of a
minimum ionizing particle.

The HERA beam gas interactions occur at c.m.\ energies similar to those
of $p\rightarrow n$ data measured at Fermilab and
the ISR~\cite{experiments} where neutron spectra were found
to be in good agreement with the predictions of
one pion exchange~\cite{bishari,experiments}. The energy scale
of FNC II was determined by fitting the observed beam gas spectrum
above 600 GeV to that expected from one pion exchange, folded
with the response of FNC II as simulated by MC.
The error in the energy scale is estimated
as 5\%.

Proton beam gas data taken
during a special run at proton 
energies of 150, 300, 448, 560, 677, and 820 GeV
showed that the energy response of FNC II was linear 
to within 4\%.

To correct for the drift in gains of the PMTs, proton beam gas
data were taken with an FNC trigger approximately every two weeks.
The mean response of each tower showed variations between
calibration runs at the level of 3\%.

The overall acceptance for neutrons, $A_{\rm FNC}$, is independent 
of the acceptance of the main detector.
To determine
$A_{\rm FNC}$,
the inactive material obscuring the aperture 
had to be modeled. About half of the inactive material was of
simple geometric shape 
and included in the ZEUS detector simulation. 
The remainder, consisting mostly of iron 
between the beam line
elements and FNC II, was  modeled by an
iron plate.
The thickness of
this plate was adjusted so that the resulting MC energy spectrum 
of neutrons from beam gas interactions matched 
the observed spectrum.
Since the interaction of neutrons in the material leads,
in general, to the loss of energy either by absorption
and/or by particle emission outside 
the acceptance of FNC II,
the observed energy spectrum is very
sensitive to the amount of inactive material upstream.
Therefore, in this study of inactive material,
events in which the neutrons began showering
upstream of FNC II were included in the spectrum; that is,
no cut was made on charged particles in the scintillator
counters in front of FNC II. The resulting thickness 
of the plate was 16$\pm$7 cm. 
Because of interactions in the inactive material, only about 15\% 
of neutrons with energy $E_n > 250$~GeV
which pass through the geometric aperture reach FNC II and
survive the scintillator cuts.
The acceptance is constant within 15\% 
for neutrons with energy $400 < E_n < 820$~GeV scattered at a fixed  
angle in the range 
0~to~0.7~mrad.

The overall acceptance assuming one pion exchange with the form
described in Section~\ref{mc} is
$4.9^{+3.0}_{-1.9}$\% 
for neutrons with $E_n>400$~GeV and $|t|<0.5$~GeV$^2$.
The error quoted is dominated by the systematic error in
estimating the amount of inactive material in front of FNC II.

To study the effect of uncertainties in the 
theoretical form of the cross section
for one pion exchange, the part of the acceptance
due to the geometric aperture, 
as shown in Fig.~\ref{calfig}(d), was calculated
for several proposed forms~\cite{kopeliovich, bishari,holtmann}.
It was found to vary from approximately 32\% to 35\%. 
This part of the acceptance for $\rho$ exchange varies between 10\%
and 30\%, 
depending on the model~\cite{kopeliovich,szczurek}.


\section{Triggering and data selection}

The selection was almost identical to that used for the 
measurement of the structure function $F_2$~\cite{z_shift}.


Events were filtered online by a three level trigger 
system~\cite{b:Detector}. 
At the first level DIS events were selected by requiring a minimum
energy deposition in the electromagnetic section of the CAL.
The threshold depended on the position in the CAL and varied between
3.4 and 4.8 GeV.
For events selected with the analysis cuts listed below, this
trigger was  more than 
99\% efficient for positrons with energy greater than 7~GeV,
as determined by Monte Carlo studies.

At the second level trigger (SLT), 
background was further reduced using the measured 
times of energy deposits and the summed energies from the calorimeter. 
The events were accepted if
\[
  \delta_{SLT} \equiv \sum_i E_i(1-\cos\theta_i) > 24
\:\:{\rm GeV} - 2E_{\gamma},
\]
where $E_i$ and $\theta_i$ are the energies and polar angles (with respect
to the primary vertex position) of calorimeter cells, and $E_{\gamma}$
is the energy deposit measured in the LUMI photon calorimeter.
For perfect detector resolution and acceptance, 
$\delta_{SLT}$ is twice the positron beam energy (55~GeV) 
for DIS events, while for photoproduction events,
where the scattered positron escapes down the beam pipe,
$\delta_{SLT}$ peaks at much lower values.

The full event information 
was available at the third level trigger (TLT).
Tighter timing cuts as 
well as algorithms to remove beam halo muons and cosmic muons were
applied.
The quantity $\delta_{TLT}$ was determined in the same manner as for
$\delta_{SLT}$. The events were required to have
$\delta_{TLT} > 25 \:\:{\rm GeV} - 2E_{\gamma}$.
Finally, events were accepted as DIS candidates if
a scattered positron candidate of energy greater than 4~GeV was found.

In the analysis of the resulting data set, further selection
criteria were applied both
to ensure accurate reconstruction of the kinematical
variables, and to increase the purity of the sample by
eliminating background from photoproduction. These cuts were:
\begin{eqnarray}
E'_e > 8\ {\rm GeV}, \nonumber\\
\yjb > 0.04, \ y_e<   0.95,  \nonumber\\
|X|  >  14\ {\rm cm}\ {\rm or}\ |Y|  >  13\ {\rm cm}, \nonumber\\
-40  <  Z_{\rm vertex}  <  40\ {\rm cm}, \nonumber\\
35  <  \delta  <  65\ {\rm GeV}, \nonumber
\end{eqnarray}
where
$y_e$ is $y$ evaluated from the scattered positron energy, $E'_e$,
and angle; 
$X$ and $Y$ are the impact position of the 
positron on the CAL as determined using the SRTD.
The cut on $|X|,|Y|$ is a fiducial volume cut to avoid
the region directly adjacent to the rear beam pipe.

Beam conditions sometimes resulted in a large FNC II
counting rate from energy deposits above the threshold
of 250 GeV.
Runs were rejected if the counting rate,
averaged over the run, was greater than 5 kHz
in order to reduce the probability of a beam gas
interaction randomly overlapping a true DIS event.
Neutron tagged events were selected by requiring that
FNC~II show an energy deposit above threshold, and that
the scintillation veto counters show an energy deposit
below that of a minimum ionizing particle.

This study is restricted
to events with $Q^2>10$ GeV$^2$~\cite{z_rap_gap}. 
After these selections, 112k events remain containing
669 neutron tagged events constituting 0.6\% of the sample.

\section{Backgrounds}

The counting rate of FNC II is predominantly due to protons
interacting with residual gas in the beam pipe.
As a result,
the main background is due to the random overlap of 
energetic neutrons from beam gas interactions
with genuine DIS events.

The fraction of beam gas triggers which survive the
scintillation counter charged particle veto
was measured to be 54$\pm 4$\%.
The average raw counting rate of FNC II during the taking of
$ep$ data was 1.5 kHz leaving
an effective counting rate of 833 Hz after the cuts. 
With 170 proton bunches in
220 HERA RF buckets and a crossing time of 96 ns, the 
overlap probability of a neutron with a random bunch was
$1.0\cdot 10^{-4}$.
Since neutrons are tagged in 0.6\%
of the events,
\[
\frac{{\rm signal}}{{\rm background}} = 
                \frac{0.6\cdot 10^{-2}}{1.0\cdot 10^{-4}} 
     = 60.
\]
Thus only 1.7\% of the neutron tagged events result from random overlaps.
The same result is obtained if the background is calculated 
on a run by run basis.

The small random coincidence rate was confirmed by the rate of
neutrons in non $ep$ background events (cosmic rays and
beam halo muons), and in a sample of random triggers.

For the DIS selection, the background from photoproduction
was estimated to be less than 1\% overall.
A sample of photoproduction events was studied
to rule out the possibility that the observed
rate of neutrons
in DIS was due to an anomalously large production rate
of neutrons in photoproduction. A fractional rate
in photoproduction comparable to
that in DIS was found, verifying 
that the photoproduction background after the neutron tag was
also less than 1\%. 
The same conclusion holds for the background
from beam gas interactions.
 
\section{Characteristics of events with a leading neutron}
\label{properties}
 
\begin{figure}[thb] 
\centerline{ \epsffile{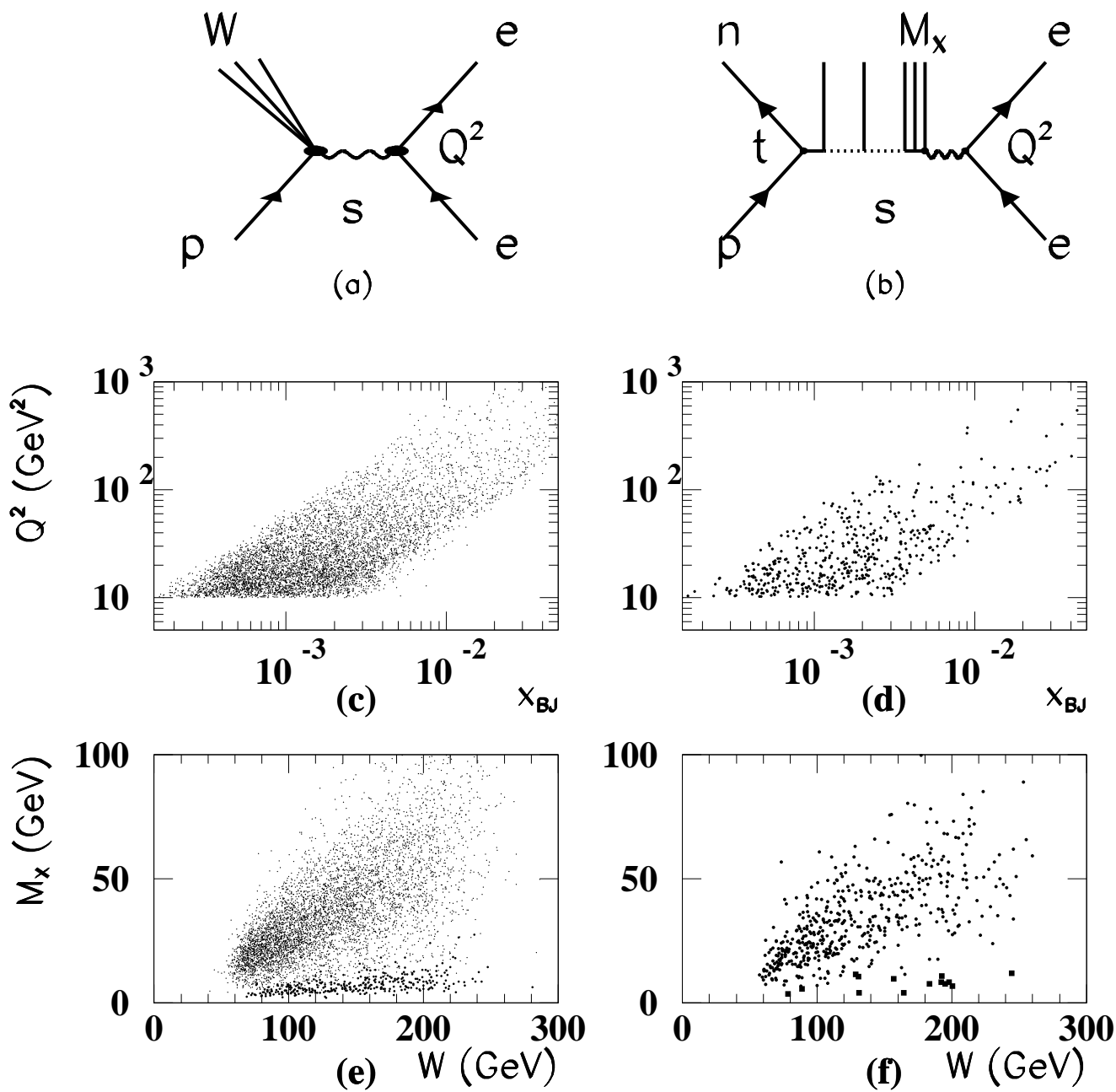} }
\vspace{1cm}
\caption{(a) Diagram for the
inclusive reaction $ep\rightarrow e+$anything, (b)  
for the two particle inclusive 
reaction $ep\rightarrow en +$anything,
a special case of (a) where the hadronic system of mass $W$ 
contains a forward neutron. The part of the hadronic system detected by
CAL is denoted by X and has a mass $\mx$.
(c) A scatter plot of $Q^2$ versus $\xbj$ for DIS events,
and (d) neutron tagged DIS events with $E_n > 400$~GeV corresponding to (c). 
(e) A scatter plot of  $\mx$ versus $W$ for DIS events.
The events in the band at low $\mx$ (larger dots) are 
the large rapidity gap events. 
(f) A scatter plot of $\mx$ versus $W$  for neutron tagged DIS events
with $E_n > 400$~GeV. The LRG events are plotted as squares.}
\label{plot1}
\end{figure}

\begin{figure}[htb] 
\centerline{ \epsffile{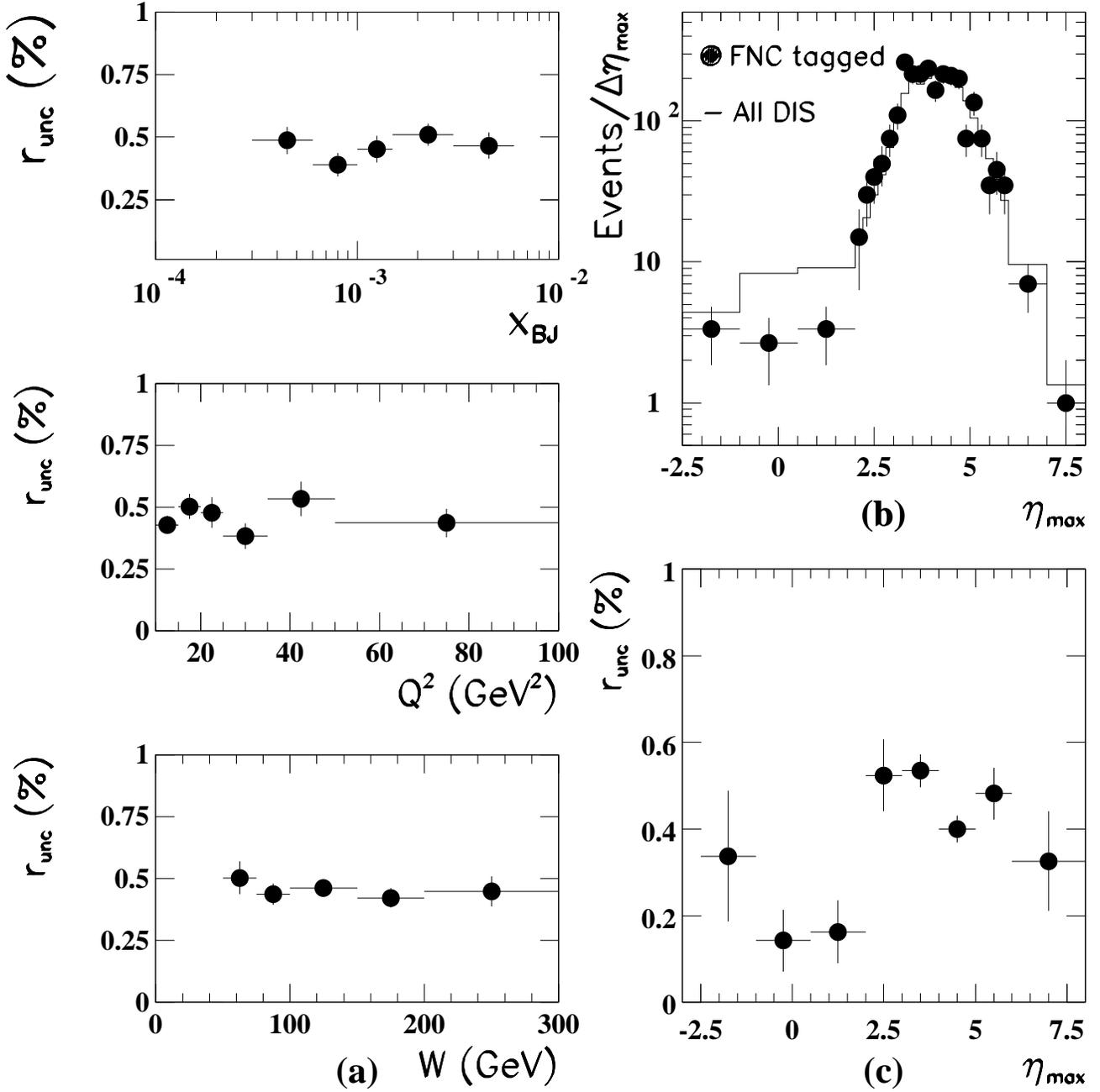} }
\caption{(a) The observed ratio of neutron tagged DIS events
with $E_n > 400$~GeV to
all DIS events as a function of $\xbj$, $Q^2$, and $W$.
(b) The data points show the $\etamax$ distribution
for tagged DIS events with $E_n > 400$~GeV. 
The distribution for all DIS events multiplied by $0.45\cdot 10^{-2}$ 
is superimposed as a histogram. 
(c) The observed ratio of tagged DIS events  with $E_n > 400$~GeV to
all DIS events as a function of $\etamax$.}
\label{plot2}
\end{figure}

The production of  neutron tagged events  with neutron energy $E_n >
400$~GeV was studied as a
function of the lepton kinematical variables.
Figure~\ref{plot1}(c) shows a scatter plot of $Q^2$ versus $\xbj$ 
for a sample of 10k DIS events which were not required to have 
a neutron tag.
All events in the full sample with a neutron tag 
are shown in
Fig.~\ref{plot1}(d).
The neutron tagged events 
follow the distribution of DIS events.
This is demonstrated quantitatively in Fig.~\ref{plot2}(a)
which shows the ratio $\runc$ 
of tagged events to all events, uncorrected for acceptance,
as a function of $\xbj$, $Q^2$ and $W$. 
Within the statistical accuracy, 
$\runc$ is consistent with being constant.
This is also true if we take the ratio as a function of
$Q^2$ in bins of $\xbj$ (not shown). Averaged over the $\xbj$ 
and $Q^2$ region the value of the ratio is 
$\rbar = 0.45\pm 0.02 \pm 0.02$~\%
for $E_n > 400$~GeV. The first error is statistical and the second
systematic. The latter is  dominated by the neutron 
energy scale uncertainty.

Further insight is gained by examining the scatter plot of
$\mx$ versus $W$ shown in Fig.~\ref{plot1}(e) for
the sample of 10k events. 
In this plot, there is a concentration of events at low $\mx$.
These events are found to have a large rapidity gap (LRG),
$\etamax < 2.0$. 
The neutron tagged events are distributed similarly to the
full sample, as seen in Fig.~\ref{plot1}(f). There is
a concentration of a few events with a rapidity gap
at low $\mx$, but most neutron tagged events are
above the low $\mx$ band.

The $\etamax$ distributions for all DIS events and for 
neutron tagged
DIS events are similar in shape
for $\etamax \gap 2$ (Fig.~\ref{plot2}(b)),
showing an exponential rise for  
$2 \lap \etamax \lap 3.5$.
Note that for $ \etamax \gap 4$ the
distributions are strongly affected by limited acceptance
towards the forward beam hole\footnote{Values of $\etamax > 4.3$
are an artifact of the clustering algorithm, and may occur
when particles are distributed in contiguous cells around
the beam pipe.}.

For $\etamax \lap 2.0 $ 
there are relatively
fewer neutron tags in the LRG events by a factor of about 2:
the small $\etamax$ events 
represent 7\% of all DIS events, but only
3\% of the neutron tagged
DIS events. This is shown in the plot of $\runc$
as a function of $\etamax$ in Fig.~\ref{plot2}(c).
LRG events with a leading neutron are expected, for instance, 
from diffractive production of a baryonic system decaying
to an energetic forward neutron and 
from double peripheral processes, where a pomeron is exchanged
between the virtual photon and the virtual pion emitted from the
proton. This effect warrants further study.

\begin{figure}[htb] 
\centerline{ \epsffile{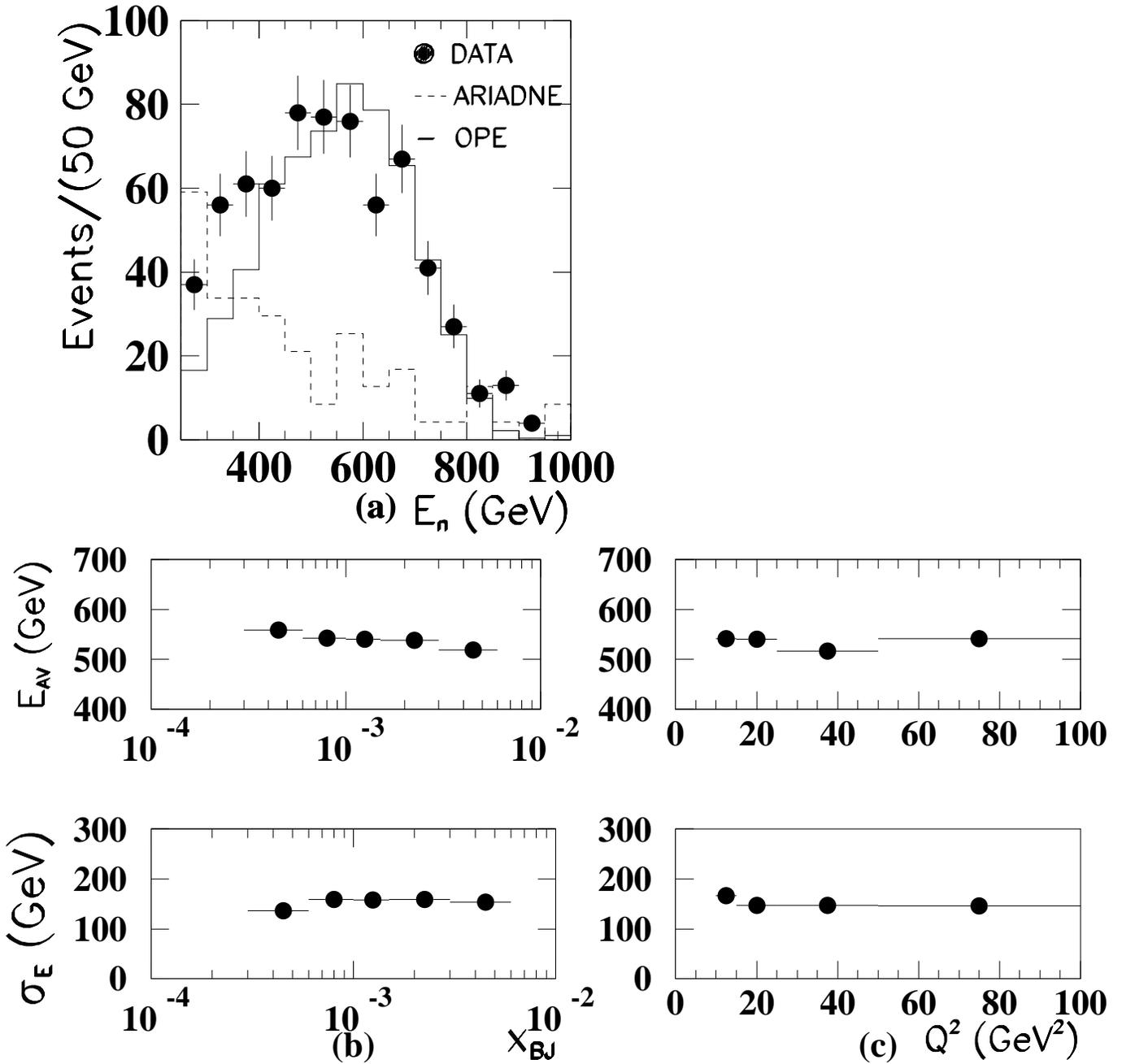} }
\caption{(a) The energy distribution of neutrons tagged
by FNC II, uncorrected for acceptance. The solid points 
are data and the histogram is the result of a 
one pion exchange DIS Monte Carlo calculation normalized
to the number of events greater than 400 GeV.
The dashed histogram gives the prediction of ARIADNE normalized
to the same luminosity as the data.
(b) and (c) The variation of the mean $E_{\rm AV}$ 
and width $\sigma_E$ of the neutron
energy spectrum above 250~GeV as a function of $\xbj$ and $Q^2$. 
}
\label{plot3} 
\end{figure}

The measured fraction of DIS events with a leading neutron
with $E_n > 400 $~GeV,  
$\rbar = 0.45\pm 0.02\pm 0.02$~\%, can be compared with the predictions
of models for DIS at HERA. 
ARIADNE~\cite{ariadne}, which is 
a colour dipole model including the boson gluon fusion
process, in general gives a good description
of the hadronic final state in DIS at HERA.
The value of $\rbar$ predicted by ARIADNE is
$0.13 \pm 0.05$\%, where the error is due to the
uncertainty in the acceptance. This is a factor
of about 3 less than that observed.
Figure~\ref{plot3}(a) shows the observed energy
spectrum of neutrons tagged above 250~GeV by FNC II.
The shape of the
neutron energy distribution predicted by ARIADNE fails to describe
the data, as seen from the dashed histogram in Fig.~4(a). 
The DIS models MEPS~\cite{meps}
and HERWIG~\cite{herwig} 
predict a higher rate of neutrons by about a factor of 2
but still fail to reproduce the observed energy spectrum.


The result of the one pion exchange 
Monte Carlo calculation of the
expected spectrum is superimposed on the energy spectrum in
Fig.~4(a), normalized to the total number
of events above 400 GeV. 
There is reasonably good agreement between 
the Monte Carlo simulation and the data at energies above 400 GeV.
At lower energies, other exchanges, such as the $\rho$, may become
important.
The neutron energy distribution 
shows no indication of
varying with $\xbj$ or $Q^2$. 
This is demonstrated in Fig.~\ref{plot3}(b) and (c) 
where the mean $E_{\rm AV}$ and width
$\sigma_E$ of the neutron energy distribution
are shown as functions of $\xbj$ and $Q^2$.

If $A_{\rm FNC}$ as determined for
one pion exchange is taken together with $\rbar$ as
measured in the data,
$9.1^{+3.6}_{-5.7}$\% of DIS events have 
a neutron with
energy $E_n>400$ GeV and $|t|<0.5$ GeV$^2$. 
The prescription
for absorptive corrections discussed in section~4
decreases this fraction by about 8\%.

\section{Conclusions}
 
We have observed energetic forward neutron 
production in DIS at HERA.
The neutrons are detected at very small
scattering angles, $\theta \lap 0.75$ mrad, and 
at high $\xl \equiv E_n/E_p$, $\xl > 0.3$.
Within present statistics leading neutron production 
is a constant fraction of DIS independent of $\xbj$ and
$Q^2$ in the range $ 3 \cdot 10^{-4} < \xbj < 6\cdot 10^{-3}$
and $10 < Q^2 < 100$~GeV$^2$. Furthermore, 
the neutron energy spectrum shows no 
variation of its mean or width 
with $\xbj$ and $Q^2$.
Neutrons with energy 
$E_n > 400$ GeV and $|t|<$ 0.5 GeV$^2$
account for a substantial fraction (at the level of 10\%) of DIS events.




\section*{Acknowledgments} 
 
We acknowledge helpful discussions with E. Gotsman, G. Ingelman,
N. Nikolaev, F. Schrempp, A. Szczurek and
P. Zerwas.
We thank F.\ Czempik, A. Kiang, H. Schult, V.\ Sturm, and K.\ Westphal
for their help with the design
and construction of the calorimeter. 
We also thank the HERA machine staff for their
forbearance during the operation of 
the FNC. We especially appreciate the
strong support provided by the DESY Directorate.
 

\newpage
\pagestyle{empty}

\newpage

\end{document}